\documentclass[pre,twocolumn]{revtex4}

\usepackage{graphicx}
\usepackage{pstricks}
\usepackage{pst-node}
\usepackage{amsmath}
\usepackage{bbold}
\usepackage{epsfig}
\usepackage{algorithm}
\usepackage{algorithmic}

\begin{document}

\title{
  Optimized network clustering by jumping sub-optimal dendrograms
}

\date{\today}

\author{Nicolas Bock}
\email{nbock@lanl.gov}
\affiliation{
  Theoretical Division, Los Alamos National Laboratory, Los Alamos, NM 87545,
  USA
}

\author{Erik Holmstr\"{o}m}
\affiliation{
  Theoretical Division, Los Alamos National Laboratory, Los Alamos, NM 87545,
  USA
}
\affiliation{
  Instituto de F\'{i}sica, Universidad Austral de Chile, Casilla 567, Valdivia,
  Chile
}

\author{Johan Br\"{a}nnlund}
\affiliation{
  Department of Mathematics \& Statistics,
  Dalhousie University,
  Halifax, NS B3H 3J5
  Canada
}

\begin{abstract}

  We propose a method to improve community division techniques in networks that
  are based on agglomeration by introducing dendrogram jumping.  The method is
  based on iterations of sub-optimal dendrograms instead of optimization of each
  agglomeration step.  We find the algorithm to exhibit excellent scaling
  behavior of its computational complexity. In its present form the algorithm
  scales as $\mathcal{O} (N^{2})$, but by using more efficient data structures
  it is possible to achieve a scaling of $\mathcal{O} (N \log^{2} N)$. We
  compare our results with other methods such as the greedy algorithm and the
  extremal optimization method. We find modularity values larger than the greedy
  algorithm and values comparable to the extremal optimization method.

\end{abstract}

\maketitle

\section{Introduction}

The study of communities in networks has received considerable attention
recently. Generally, a community can be thought of as a subset of nodes of the
network in which the nodes within a community are more connected among each
other than they are connected to the other nodes in the network.  By analyzing a
network in terms of its communities, it is possible to gain understanding of the
structure of a network on a larger scale and to uncover previously unnoticed
connections between nodes or groups of nodes.  Examples of successful community
division studies include a study on the relationship between diseases and genes
\cite{Wilkinson2004}, the identification of transition states in potential
energy landscapes \cite{MassenCP:Idecwe}, and the identification of recording
locations and racial community structures in a jazz musicians network in the USA
around 1920 \cite{GleiserPM:Comsj}. We have recently used our modularity
optimization algorithm to optimize the performance of a recursive inverse
factorization technique used in large scale electronic structure calculations
\cite{Rubensson}.

The analysis of a network in terms of communities poses a difficult challenge.
The clustering algorithm has to be accurate so that it identifies informative
community structures.  This implies that the algorithm has to consider many of
the possible community divisions in the network before it can decide which one
is the best.  From a computational point of view we are faced with a rapidly
growing problem as a function of network size. To consider all possible
community divisions becomes computationally unfeasible even for relatively small
networks. In fact, it can be shown that the number of ways of dividing a network
into communities grows as the Sterling number of the second kind
\cite{NewmanMEJ:Fasadc}.

In order to quantify the quality of a community division, a quality function is
introduced that assigns a ``value-reflecting'' number, or ``quality-of-split''
\cite{Ward63} to a community division. The modularity, as introduced by
\citet{GirvanM:Comssa}, is a popular choice for such a quality function.
Although there may be some drawbacks to this approach as pointed out by
\citet{Fortunato}, the community divisions that are obtained by optimizing the
modularity typically give valuable information \cite{Wilkinson2004,
MassenCP:Idecwe}.

In the literature many different optimization strategies can be found that
employ the modularity. They vary in quality, i.e.\ the value of the largest
modularity they find, and in the computational effort. An efficient
agglomerative method is the greedy algorithm of \citet{NewmanMEJ:Fasadc} which
\citet{ClausetAaron:Fincsv} showed to run in $\mathcal{O} (N \log^{2} N)$
computational effort. Other methods however, find modularity values larger than
the greedy algorithm. These include extremal optimization \cite{DuchJ.:Comdcn},
basin-hopping \cite{MassenCP:Idecwe}, simulated annealing
\cite{DorsoCO:Detcsn}, recursive filtration \cite{Shen20086663}, a heuristic
algorithm \cite{Chen20092741}, and a spectral algorithm
\cite{NewmanMEJ:Modacs}. 

In this article we introduce a new agglomerative method which we demonstrate by
using the modularity as the quality-of-split function. Our method is general
however and can be used with any other quality-of-split function. We find
modularity values comparable to the extremal optimization method for a set of
well-studied networks. In section~\ref{sec:Theory} we summarize the
agglomerative greedy algorithm and introduce our method. In
section~\ref{sec:Results} we compare with results for some popular networks
using the greedy method and the extremal optimization method.  Finally, in
section~\ref{sec:Conclusions} we present our conclusions.

\section{Theory}
\label{sec:Theory}

A network of $N$ nodes can be divided into any number of communities, $C$, where
$1 \leq C \leq N$. The extremal cases $C = 1$ and $C = N$ are the two trivial
solutions in which all nodes either belong to only one common community or each
belongs only to its own separate community, respectively. Given that the number
of possible community divisions is exceedingly large for any decently sized
network, the problem of finding the optimal community split cannot be approached
by calculating all possible splits. Instead one has to resort to approximate
solutions of the problem. One possibility is to attempt to find the optimal
community division by starting with one of the two trivial cases and proceeding
by either stepwise merging two communities or by splitting a community into two
until the opposite extremal case is reached. These two approaches are commonly
referred to as agglomerative and divisive methods, respectively. The set of
community divisions for every value of $C$ between $1 \leq C \leq N$ is called a
dendrogram.  Such dendrogram-based optimization methods aim to find the best
community split by optimizing each step along the dendrogram based on the change
in the quality-of-split measure \cite{NewmanMEJ:Fasadc, GustafssonM:Comavc,
NewmanMEJ:Finaec}.  These methods are typically very efficient computationally,
but the quality of their result depends critically on the heuristic chosen
during the stepwise optimization process. In fact, in a recent comparison of
dendrogram-based methods and a simulated annealing technique,
\citet{DanonL.:Comcsi} found that the simulated annealing method, which is not
bound to a dendrogram, is able to find better solutions to the community
division problem than the dendrogram-based methods.

The quality-of-split function we will use for this study is the modularity, $Q$,
of \citet{GirvanM:Comssa}. It is defined as 

\begin{equation}
  Q = \mathrm{Tr} (e) - \sum_{ij} (e^2)_{ij},
  \label{eq:modularity}
\end{equation}

\noindent
where $e$ is the assortative mixing matrix (which is a $C \times C$ matrix).
The elements $e_{ij}$ are given by the number of links from community $i$ to $j$
for a particular community division as a fraction of the total number of links.

The greedy algorithm was introduced by Newman \cite{NewmanMEJ:Fasadc} and is a
good example of an agglomerative method. A range of other agglomerative methods
have been proposed which differ in how each step in the merge process is chosen
\cite{NewmanMEJ:Fasadc, Ward63, MacQueen67, ClausetAaron:Fincsv}. The change in
modularity due to merging two communities $i$ and $j$ can be written as
\cite{NewmanMEJ:Fasadc}

\begin{equation}
  \Delta Q = e_{ij} + e_{ji} - a_{i} b_{j} - a_{j} b_{i},
  \label{eq:deltaQ}
\end{equation}

\noindent
where $a_{i}$ and $b_{i}$ are the column and row sums of $e$, respectively. Each
merge process is chosen such as to maximize the effected modularity change in
the hope that this leads to the community division with the maximum modularity.
What makes the greedy method particularly attractive is the fact that
eq.~(\ref{eq:deltaQ}) is inexpensive to evaluate (it is of $\mathcal{O} (1)$
computational effort). In addition, the number of community merges to evaluate
is at most $(N - 1) (N - 2) \cdots$ which leads to an overall computational
complexity of $\mathcal{O} (N \log^{2} N)$ \cite{ClausetAaron:Fincsv}.

\subsection{Sub-optimal iterations}

Optimization techniques which are bound to a dendrogram generally find smaller
values of the modularity than methods which do not operate along a dendrogram
\cite{MedusA:Detcsn}. This is due to the fact that the optimal solution may not
be accessible by walking a stepwise optimized dendrogram.  Optimal choices in
the beginning of the dendrogram (large $C$ for agglomerative methods) may lead
to community divisions which can not be merged further to achieve the optimal
division. This is of course also a problem for divisive methods. It may
therefore not be possible to find a heuristic that finds the optimal community
division along one single dendrogram.  Motivated by our previous work on the
modularity density of networks \cite{Holmstrom20091161} we avoid the problem
by performing incomplete optimizations in each merge process. We do not
consider all $C (C-1)$ possible merges, but only a smaller randomly chosen
subset and pick the merge with the largest modularity increase from this
subset.  This procedure has the advantage that it allows for randomness in the
agglomeration process so that two different runs will not give the same
result.  The randomness is obviously also a drawback since it is very unlikely
that we find the optimal community split in one such random dendrogram. Even
several sub-optimal dendrograms will be unlikely to have produced the optimal
community division. We therefore iterate over several random dendrograms in an
inner loop and optimize a list of modularity values for each value of $C$ in
an outer loop.  Our algorithm is expressed in terms of a pseudocode in
figure~\ref{alg:dendJump}.

\begin{figure}
  \begin{algorithmic}[1]
    \FOR[Outer loop]{$N^{\mathrm{outer}}$ times}
      \FOR[Inner loop]{$N^{\mathrm{inner}}$ times}
        \FOR[Suboptimal agglomeration]{$C = N \mbox{ to } C \geq 2$}
          \STATE Find $n$ random merge candidates.
          \STATE Calculate $\Delta Q_{n} (C \rightarrow C-1)$ from eq.~(\ref{eq:deltaQ}).
          \IF[Accept proposed merge process]{$\max (Q^{\mathrm{inner}}_{C} + \Delta Q_{n}) > Q^{\mathrm{inner}}_{C-1}$}
            \STATE $e^{\mathrm{inner}}_{C} \Leftarrow e$
            \STATE $Q^{\mathrm{inner}}_{C-1} \Leftarrow Q^{\mathrm{inner}}_{C} + \Delta Q_{n}$
            \STATE Calculate new $e$.
          \ELSE[Reject proposed merge process]
            \STATE $e \Leftarrow e^{\mathrm{inner}}_{C-1}$
          \ENDIF
        \ENDFOR \COMMENT{Suboptimal agglomeration}
      \ENDFOR \COMMENT{Inner loop}
      \FOR[Update list of modularity values]{$C = 1 \mbox{ to } C < N$}
        \IF{$Q^{\mathrm{inner}}_{C} > Q^{\mathrm{outer}}_{C}$}
          \STATE $Q^{\mathrm{outer}}_{C} \Leftarrow Q^{\mathrm{inner}}_{C}$
        \ENDIF
      \ENDFOR \COMMENT{Update list of modularity values}
    \ENDFOR \COMMENT{Outer loop}
  \end{algorithmic}
  \caption{
    \label{alg:dendJump}
    Sub-Optimal Dendrogram Jumping Algorithm: (line 3) This loop performs the
    suboptimal agglomeration. In this study $Q$ means the modularity, but any
    quality-of-split function can be used. (line 2) The inner loop restarts the
    agglomeration process. We improve on the best modularities found by
    accepting proposed merge processes only if they lead to higher modularity
    values. (line 1) The best modularity values found in the inner loop are
    stored and the inner loop is restarted.
  }
\end{figure}

Inside the inner loop we analyze a ``somewhat'' random dendrogram. By this we
mean that we pick the best out of $n$ randomly chosen possible merge processes
in each step of the dendrogram (line 4 in figure~\ref{alg:dendJump}).  Clearly,
in the limit $n \rightarrow 1$, the dendrogram is random. The greedy method, on
the other hand, corresponds to at most $n = C (C-1)$ different merge processes.
For any value $n < C (C-1)$ we achieve sub-optimal agglomeration. As we
decrement $C$ and walk the dendrogram from larger to smaller $C$ values, the
merge process with the largest modularity gain becomes the proposed merge
process. We store a list of the best modularities found so far for each value of
$C$ and merge along the proposed merge process only if the proposed merge
produces a higher modularity value than the previous best value.  If the
proposed merge process produces a modularity value equal or less than the best
modularity value thus far, we discard it and load the community division
corresponding to the best modularity found so far to continue. It is this step
(lines 6 through 12 in figure~\ref{alg:dendJump}) which leads to a dendrogram
jump.  The list of best modularity values of the inner loop usually converges
rather quickly and the algorithm cannot improve on the modularities any longer.

In the outer loop we store another list of best modularity values. Once the
inner loop completes we update the list of the outer loop with the values found
in the inner loop (lines 15 through 19 in figure~\ref{alg:dendJump}).  We then
reinitialize the inner loop for another run. This step allows the algorithm to
explore a different part of the community division space since it will randomly
choose other dendrograms.

The 3 tunable parameters in our algorithm, $n$, $N^{\mathrm{inner}}$, and
$N^{\mathrm{outer}}$, are chosen according to the following heuristic: (1) The
number of suboptimal trials, $n$, should be significantly smaller than the
greedy limit, $C (C-1)$. Although our method will work with any $n \ge 1$, we
have found in practice that any value between $5 \le n \le 10$ works well. (2)
The number of inner loops, $N^{\mathrm{inner}}$, should be chosen as small as
possible such that the $Q_{C}^{\mathrm{inner}}$ list is converged. This process
is exemplified by the convergence of the blue dashed lines in the upper panel of
figure~\ref{fig:GenDendrogram}. The value of this particular parameter depends
strongly on the network size. We have found values of $N^{\mathrm{inner}}$
between 8 and 150 for the smaller networks (Zachary Karate Club) and the larger
networks (Jazz musician and e--mail network), respectively, to work well. (3)
There is no predetermined limit on the number of outer loops,
$N^{\mathrm{outer}}$. It is determined by the level of convergence desired for
the maximum modularity value. The more outer loops are performed, the more
likely it is to find the maximum modularity value.

Iterating over the inner and outer loop converges rapidly and we find a list of
the best modularity values for all $C$. In the following section we will compare
our results with previously studied networks.

\section{Results}
\label{sec:Results}

\begin{figure}
  \begin{tabular}{r}
    \includegraphics[width=0.90\linewidth]{ZacharysKarateClub.Unw.QMax.4.eps}
    \\
    \includegraphics[width=0.85\linewidth]{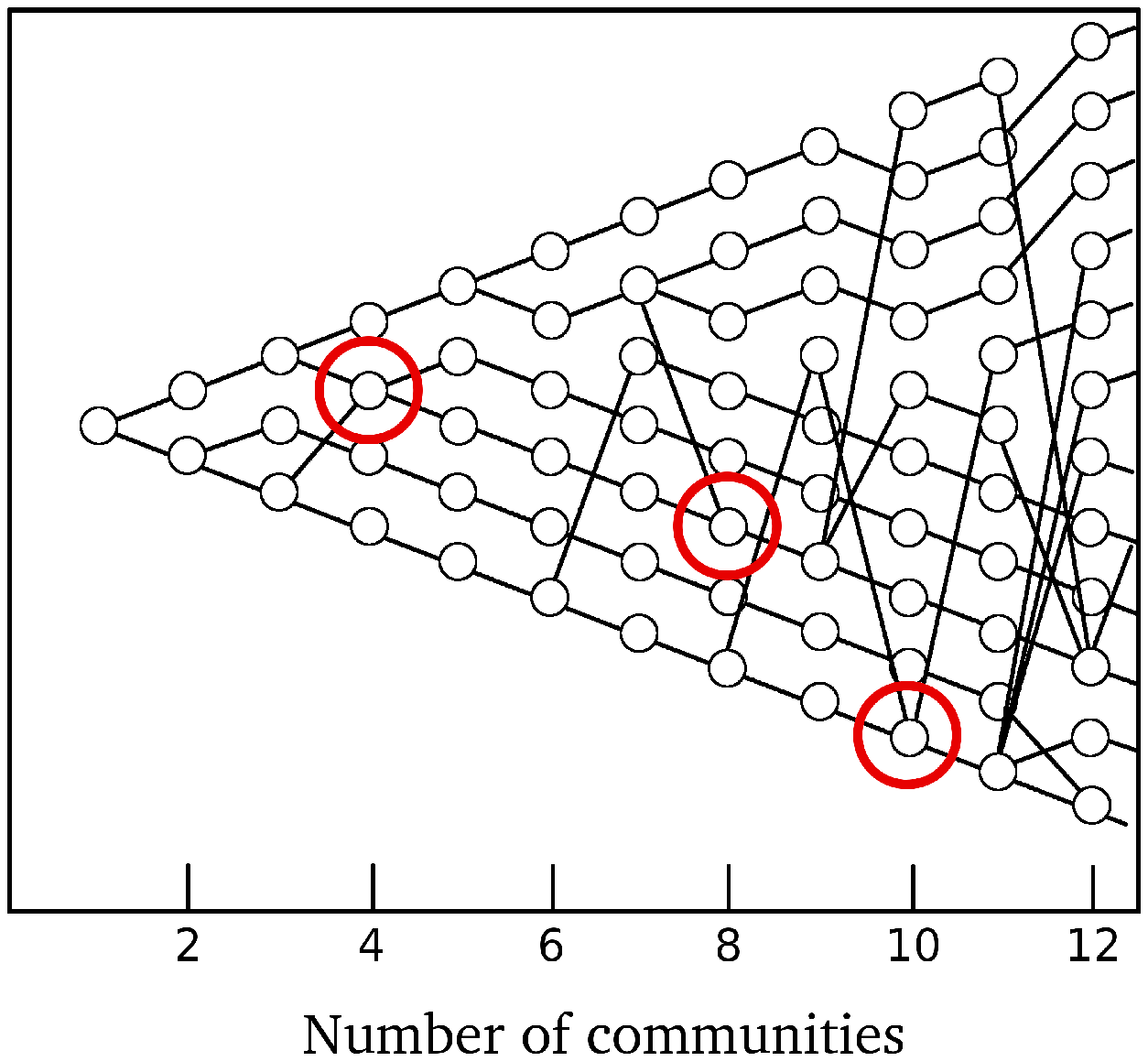}
  \end{tabular}
  \caption{
    {\bf Upper panel}: 
    The dashed lines (blue online) indicate the maximum modularity values found
    in the inner loop. With each iteration of the inner loop the maximum
    modularity increases and eventually converges, shown with the bold dashed
    line (blue online). The circles show the maximum modularity values found
    after the outer loop is converged. The vertical lines indicate sections that
    belong to one dendrogram. Dendrogram jumping occurs across the vertical
    lines.
    {\bf Lower panel}: 
    A generalized dendrogram that corresponds to the maximum modularity of the
    unweighted Zachary Karate Club network. Events that indicate dendrogram
    jumps are marked with red circles.  
  }
	\label{fig:GenDendrogram}
\end{figure}

In the upper panel of figure~\ref{fig:GenDendrogram} we show the best modularity
values found by our iterative dendrogram jumping method for the unweighted
Zachary Karate Club network (black circles). This result was obtained with
$N^{\mathrm{inner}} = 8$ inner loop iterations and $N^{\mathrm{outer}} = 20$
outer loop iterations. The number of sub-optimal trials in each agglomeration
was $n = 10$. We indicate the borders between sections of the curve that belong
to the same dendrogram to illustrate the dendrogram jumping of our algorithm. In
the case shown, we find 12 such borders.  We find the maximum modularity at $C =
4$ at a value of $Q = 0.4198$. The blue dashed lines show the evolution of an
optimization in the inner loop consisting of 8 sub-optimized $Q(C)$ curves
obtained by 10 merge trials per $C$. We find that this particular run over the
inner loop of the dendrogram jumping method achieves high modularities for $C =$
3, 7, and 8 but fails to find the largest modularity that we find after looping
over the outer loop.

In the lower panel of figure~\ref{fig:GenDendrogram} we show the first few
branches of the resulting disconnected dendrogram for the overall optimization
of the Zachary Karate Club network. At $C = 1$, all nodes are in the same
community as indicated by the single black circle. As this community is split
into two and subsequently three communities, we find that the figure looks like
an ordinary dendrogram.  However, splitting 3 into 4 communities, the community
division corresponding to the highest modularity value results in the merging of
two communities and the splitting of two communities. We indicate this merging
by the first red circle.  This is a situation that cannot occur in a dendrogram
and thus indicates a dendrogram jump in the maximum modularity curve.  Two more
such events are marked with red circles.

We have performed modularity optimizations by means of our improved sub-optimal
dendrogram jumping algorithm for a set of well-known networks.  In
table~\ref{tab:QMax} we present our results for the maximum modularity and the
number of communities that were obtained. The results are compared to the
corresponding maximum modularity found by the greedy method
\cite{NewmanMEJ:Modacs} and the extremal optimization method
\cite{DuchJ.:Comdcn}, where applicable. We find that dendrogram jumping always
finds modularity values larger than the greedy algorithm and values comparable
to the extremal optimization method. In most cases the number of communities
corresponding to the largest modularity value differs from what was found with
the other methods. We never found a case in which the communities had the same
members in the different methods even in the cases were the number of
communities was the same. For the small networks the community members differed
only in few nodes. For the larger networks however, we found more significant
differences in the assignment of nodes to communities.

The evaluation of $\Delta Q$ of eq.~(\ref{eq:deltaQ}) can be done in
computational effort $\mathcal{O} (1)$ time. The subsequent update of $e$ after
a merge operation takes $\mathcal{O} (N)$ worst time and there are $N-1$ such
merge operations per dendrogram. We iterate through a fixed number of inner and
outer loops, i.e. the total computational effort is $\mathcal{O} (N^{2})$. Not
surprisingly, this is identical to the computational effort found for the greedy
algorithm \cite{NewmanMEJ:Fasadc} in the sparse graph limit. The inner loop of
our algorithm is a generalization of the greedy method in the limit of $n
\rightarrow C (C-1)$. In our algorithm we only consider a small fixed number $n$
of merge candidates, which implies that our method always scales $\mathcal{O}
(N^{2})$ even in the dense graph case.  However the data structures used by
\citet{ClausetAaron:Fincsv} to speed up the greedy algorithm clearly are readily
applicable in our case as well, which makes it possible to reduce the
computational effort to $\mathcal{O} (N \log^{2} N)$.  For comparison, the
extremal optimization \cite{DuchJ.:Comdcn} method runs in $\mathcal{O} (N^{2}
\log N)$ time.

We have successfully used our algorithm in its current implementation for
networks of up to $\approx 1200$ nodes but networks an order of magnitude larger
should be analyzable with a current desktop workstation and sufficient memory
installed.

\begin{table*}
\centering
\begin{tabular}{lcrrcccccc}
\hline 
\hline 
\  Network \  & \ Ref. \ & \ Nodes \ & \ Edges \ & \ $Q^{\mathrm{max}}$ \ & \ $C$  \ & \
$Q^{\mathrm{greedy}}$ \ & \ $C^{\mathrm{greedy}}$ \ & \ $Q^{\mathrm{EO}}$ \ & \
$C^{\mathrm{EO}}$\\
\hline
Zachary                & \onlinecite{ZacharyURL,Zachary77}           & 34   & 156   & 0.4198 & 4  & 0.3807  & 3  & 0.4188 \ & \ 4 \\
Zachary (W)            & \onlinecite{ZacharyURL,Zachary77}           & 34   & 156   & 0.4449 & 4  & 0.4345  & 3  \\
Fraternities Subj. (W) & \onlinecite{ZacharyURL,Bernard80,Bernard82} & 58   & 3306  & 0.0486 & 3  & 0.0412  & 3  \\
Fraternities Obj. (W)  & \onlinecite{ZacharyURL,Bernard80,Bernard82} & 58   & 1934  & 0.1460 & 6  & 0.1408  & 6  \\
Dolphins               & \onlinecite{NewmanURL,Lusseau2003}          & 62   & 318   & 0.5285 & 5  & 0.4923  & 4  \\
Prisoners              & \onlinecite{ZacharyURL,MacRae60}            & 67   & 182   & 0.6232 & 9  & 0.6217  & 9  \\
Les Miserables (W)     & \onlinecite{ZacharyURL,Knuth93}             & 77   & 508   & 0.5667 & 6  & 0.5472  & 5   \\
Les Miserables         & \onlinecite{ZacharyURL,Knuth93}             & 77   & 508   & 0.5600 & 6  & 0.5006  & 5  \\
Grassland              & \onlinecite{FoodwebURL}                     & 88   & 274   & 0.6627 & 9  & 0.6609  & 10   \\
Jazz bands             & \onlinecite{GleiserPM:Comsj,GleiserNote}    & 198  & 5484  & 0.4450 & 4  & 0.4389  & 4  & 0.4452 &  5\\
Littlerock             & \onlinecite{FoodwebURL}                     & 183  & 4886  & 0.3629 & 4  & 0.3395  & 3  \\
Jazz musicians         & \onlinecite{GleiserPM:Comsj,GleiserNote}    & 1265 & 76714 & 0.5780 & 18 & 0.5235  & 20 \\
e-mail                 & \onlinecite{Guimera2003,GuimeraNote}        & 1133 & 10902 & 0.5718 & 11 & 0.5093  & 15  &  0.5738 & 15\\
\hline 
\hline 
\end{tabular}
\caption{
  Our results for the optimized networks in this study compared to the greedy
  algorithm and the extremal optimization method were applicable. The entries
  labeled (W) are weighted networks. Shown are the number of nodes in the
  network (Nodes), the number of directed edges (Edges), the maximum modularity
  found by our method ($Q^{\mathrm{max}}$), the number of communities for this
  value of the modularity ($C$), the same quantities for the greedy algorithm
  ($Q^{\mathrm{greedy}}$ and $C^{\mathrm{greedy}}$), as well as for the extremal
  optimization method ($Q^{\mathrm{EO}}$ and $C^{\mathrm{EO}}$).
}
\label{tab:QMax}
\end{table*}

\section{Conclusions}
\label{sec:Conclusions}

We have presented a generic optimization technique that applies to community
detection algorithms that are agglomerative. We demonstrated the efficiency of
our method by calculating the maximum modularity for a set of networks and
comparing our results with two other methods, the greedy method by Newman and
coworkers \cite{NewmanMEJ:Fasadc, ClausetAaron:Fincsv} and the extremal
optimization method by \citet{DuchJ.:Comdcn}. In this comparison we found
modularity values for all examples studied that are larger than the results of
the greedy algorithm and comparable to the results of the extremal optimization
method. The computational complexity of our method ultimately is $\mathcal{O} (N
\log^{2} N)$. Our method therefore has the same computational complexity scaling
behavior as the greedy method. The extremal optimization method is
computationally more expensive and scales with a computational complexity of
$\mathcal{O} (N^{2} \log N)$. In applications in which sufficient memory is
available, our algorithm therefore is the method of choice.

Our study showed that the community divisions of maximum modularity are not
connected by one single dendrogram and thus any method which aims at optimizing
the modularity by optimizing each step along a dendrogram will fail.  This
finding confirms a similar conclusion drawn by \citet{MedusA:Detcsn}.  This has
important implications for the further development of modularity optimization
methods.

\section{Acknowledgments}

We would like to thank Travis Peery, Eric Chisolm, and Anders Niklasson for many
helpful discussions. We would also like to thank Mark Newman who graciously
provided us with his network data which we analyzed in this paper.  Special
acknowledgments also to the genuine and professional scientific atmosphere
provided by the International Ten Bar Caf\'{e}.

\bibliography{paper}

\begin{thebibliography}{32}
\expandafter\ifx\csname natexlab\endcsname\relax\def\natexlab#1{#1}\fi
\expandafter\ifx\csname bibnamefont\endcsname\relax
  \def\bibnamefont#1{#1}\fi
\expandafter\ifx\csname bibfnamefont\endcsname\relax
  \def\bibfnamefont#1{#1}\fi
\expandafter\ifx\csname citenamefont\endcsname\relax
  \def\citenamefont#1{#1}\fi
\expandafter\ifx\csname url\endcsname\relax
  \def\url#1{\texttt{#1}}\fi
\expandafter\ifx\csname urlprefix\endcsname\relax\def\urlprefix{URL }\fi
\providecommand{\bibinfo}[2]{#2}
\providecommand{\eprint}[2][]{\url{#2}}

\bibitem[{\citenamefont{Wilkinson and Huberman}(2004)}]{Wilkinson2004}
\bibinfo{author}{\bibfnamefont{D.}~\bibnamefont{Wilkinson}} \bibnamefont{and}
  \bibinfo{author}{\bibfnamefont{B.~A.} \bibnamefont{Huberman}},
  \bibinfo{journal}{PNAS} \textbf{\bibinfo{volume}{101}}, \bibinfo{pages}{5241}
  (\bibinfo{year}{2004}).

\bibitem[{\citenamefont{Massen and Doye}(2005)}]{MassenCP:Idecwe}
\bibinfo{author}{\bibfnamefont{C.~P.} \bibnamefont{Massen}} \bibnamefont{and}
  \bibinfo{author}{\bibfnamefont{J.~P.~K.} \bibnamefont{Doye}},
  \bibinfo{journal}{Phys. Rev. E} \textbf{\bibinfo{volume}{71}},
  \bibinfo{pages}{046101 } (\bibinfo{year}{2005}).

\bibitem[{\citenamefont{Gleiser and Danon}(2003)}]{GleiserPM:Comsj}
\bibinfo{author}{\bibfnamefont{P.~M.} \bibnamefont{Gleiser}} \bibnamefont{and}
  \bibinfo{author}{\bibfnamefont{L.}~\bibnamefont{Danon}},
  \bibinfo{journal}{Advances in Complex Systems} \textbf{\bibinfo{volume}{6}},
  \bibinfo{pages}{565 } (\bibinfo{year}{2003}).

\bibitem[{\citenamefont{Rubensson et~al.}(2008)\citenamefont{Rubensson, Bock,
  Holmstr\"{o}m, and Niklasson}}]{Rubensson}
\bibinfo{author}{\bibfnamefont{E.~H.} \bibnamefont{Rubensson}},
  \bibinfo{author}{\bibfnamefont{N.}~\bibnamefont{Bock}},
  \bibinfo{author}{\bibfnamefont{E.}~\bibnamefont{Holmstr\"{o}m}},
  \bibnamefont{and} \bibinfo{author}{\bibfnamefont{A.~M.~N.}
  \bibnamefont{Niklasson}}, \bibinfo{journal}{J. Chem. Phys.}
  \textbf{\bibinfo{volume}{128}}, \bibinfo{pages}{104105}
  (\bibinfo{year}{2008}).

\bibitem[{\citenamefont{Newman}(2004)}]{NewmanMEJ:Fasadc}
\bibinfo{author}{\bibfnamefont{M.~E.~J.} \bibnamefont{Newman}},
  \bibinfo{journal}{Phys. Rev. E} \textbf{\bibinfo{volume}{69}},
  \bibinfo{pages}{066133 } (\bibinfo{year}{2004}).

\bibitem[{\citenamefont{Ward}(1963)}]{Ward63}
\bibinfo{author}{\bibfnamefont{J.~H.} \bibnamefont{Ward}},
  \bibinfo{journal}{Journal of the American Statistical Association}
  \textbf{\bibinfo{volume}{58}}, \bibinfo{pages}{236} (\bibinfo{year}{1963}).

\bibitem[{\citenamefont{Girvan and Newman}(2002)}]{GirvanM:Comssa}
\bibinfo{author}{\bibfnamefont{M.}~\bibnamefont{Girvan}} \bibnamefont{and}
  \bibinfo{author}{\bibfnamefont{M.~E.~J.} \bibnamefont{Newman}},
  \bibinfo{journal}{PNAS} \textbf{\bibinfo{volume}{99}}, \bibinfo{pages}{7821 }
  (\bibinfo{year}{2002}).

\bibitem[{\citenamefont{Fortunato and Barth{\'{e}}lemy}(2006)}]{Fortunato}
\bibinfo{author}{\bibfnamefont{S.}~\bibnamefont{Fortunato}} \bibnamefont{and}
  \bibinfo{author}{\bibfnamefont{M.}~\bibnamefont{Barth{\'{e}}lemy}},
  \bibinfo{journal}{PNAS} \textbf{\bibinfo{volume}{104}}, \bibinfo{pages}{36 }
  (\bibinfo{year}{2006}).

\bibitem[{\citenamefont{Clauset et~al.}(2004)\citenamefont{Clauset, Newman, and
  Moore}}]{ClausetAaron:Fincsv}
\bibinfo{author}{\bibfnamefont{A.}~\bibnamefont{Clauset}},
  \bibinfo{author}{\bibfnamefont{M.~E.~J.} \bibnamefont{Newman}},
  \bibnamefont{and} \bibinfo{author}{\bibfnamefont{C.}~\bibnamefont{Moore}},
  \bibinfo{journal}{Phys. Rev. E} \textbf{\bibinfo{volume}{70}},
  \bibinfo{pages}{066111} (\bibinfo{year}{2004}).

\bibitem[{\citenamefont{Duch and Arenas}(2005)}]{DuchJ.:Comdcn}
\bibinfo{author}{\bibfnamefont{J.}~\bibnamefont{Duch}} \bibnamefont{and}
  \bibinfo{author}{\bibfnamefont{A.}~\bibnamefont{Arenas}},
  \bibinfo{journal}{Phys. Rev. E} \textbf{\bibinfo{volume}{72}},
  \bibinfo{pages}{27104 } (\bibinfo{year}{2005}).

\bibitem[{\citenamefont{Dorso et~al.}(2005)\citenamefont{Dorso, Medus, and
  Acuna}}]{DorsoCO:Detcsn}
\bibinfo{author}{\bibfnamefont{C.~O.} \bibnamefont{Dorso}},
  \bibinfo{author}{\bibfnamefont{A.}~\bibnamefont{Medus}}, \bibnamefont{and}
  \bibinfo{author}{\bibfnamefont{G.}~\bibnamefont{Acuna}},
  \bibinfo{journal}{Physica A} \textbf{\bibinfo{volume}{358}},
  \bibinfo{pages}{593 } (\bibinfo{year}{2005}).

\bibitem[{\citenamefont{Shen et~al.}(2008)\citenamefont{Shen, Pei, Wang, Li,
  and Wang}}]{Shen20086663}
\bibinfo{author}{\bibfnamefont{Y.}~\bibnamefont{Shen}},
  \bibinfo{author}{\bibfnamefont{W.}~\bibnamefont{Pei}},
  \bibinfo{author}{\bibfnamefont{K.}~\bibnamefont{Wang}},
  \bibinfo{author}{\bibfnamefont{T.}~\bibnamefont{Li}}, \bibnamefont{and}
  \bibinfo{author}{\bibfnamefont{S.}~\bibnamefont{Wang}},
  \bibinfo{journal}{Physica A: Statistical Mechanics and its Applications}
  \textbf{\bibinfo{volume}{387}}, \bibinfo{pages}{6663 }
  (\bibinfo{year}{2008}).

\bibitem[{\citenamefont{Chen et~al.}(2009)\citenamefont{Chen, Fu, and
  Shang}}]{Chen20092741}
\bibinfo{author}{\bibfnamefont{D.}~\bibnamefont{Chen}},
  \bibinfo{author}{\bibfnamefont{Y.}~\bibnamefont{Fu}}, \bibnamefont{and}
  \bibinfo{author}{\bibfnamefont{M.}~\bibnamefont{Shang}},
  \bibinfo{journal}{Physica A: Statistical Mechanics and its Applications}
  \textbf{\bibinfo{volume}{388}}, \bibinfo{pages}{2741 }
  (\bibinfo{year}{2009}).

\bibitem[{\citenamefont{Newman}(2006)}]{NewmanMEJ:Modacs}
\bibinfo{author}{\bibfnamefont{M.~E.~J.} \bibnamefont{Newman}},
  \bibinfo{journal}{PNAS} \textbf{\bibinfo{volume}{103}}, \bibinfo{pages}{8577
  } (\bibinfo{year}{2006}).

\bibitem[{\citenamefont{Gustafsson et~al.}(2006)\citenamefont{Gustafsson,
  Hornquist, and Lombardi}}]{GustafssonM:Comavc}
\bibinfo{author}{\bibfnamefont{M.}~\bibnamefont{Gustafsson}},
  \bibinfo{author}{\bibfnamefont{M.}~\bibnamefont{Hornquist}},
  \bibnamefont{and} \bibinfo{author}{\bibfnamefont{A.}~\bibnamefont{Lombardi}},
  \bibinfo{journal}{Physica A} \textbf{\bibinfo{volume}{367}},
  \bibinfo{pages}{559 } (\bibinfo{year}{2006}).

\bibitem[{\citenamefont{Newman and Girvan}(2004)}]{NewmanMEJ:Finaec}
\bibinfo{author}{\bibfnamefont{M.~E.~J.} \bibnamefont{Newman}}
  \bibnamefont{and} \bibinfo{author}{\bibfnamefont{M.}~\bibnamefont{Girvan}},
  \bibinfo{journal}{Phys. Rev. E} \textbf{\bibinfo{volume}{69}},
  \bibinfo{pages}{026113 } (\bibinfo{year}{2004}).

\bibitem[{\citenamefont{Danon et~al.}(2005)\citenamefont{Danon, Diaz-Guilera,
  Duch, and Arenas}}]{DanonL.:Comcsi}
\bibinfo{author}{\bibfnamefont{L.}~\bibnamefont{Danon}},
  \bibinfo{author}{\bibfnamefont{A.}~\bibnamefont{Diaz-Guilera}},
  \bibinfo{author}{\bibfnamefont{J.}~\bibnamefont{Duch}}, \bibnamefont{and}
  \bibinfo{author}{\bibfnamefont{A.}~\bibnamefont{Arenas}},
  \bibinfo{journal}{Journal of Statistical Mechanics: Theory and Experiment}
  \textbf{\bibinfo{volume}{2005}}, \bibinfo{pages}{P09008}
  (\bibinfo{year}{2005}).

\bibitem[{\citenamefont{MacQueen}(1967)}]{MacQueen67}
\bibinfo{author}{\bibfnamefont{J.}~\bibnamefont{MacQueen}},
  \bibinfo{journal}{In: Le Cam L.M., Neyman J. (Eds.), Proc. 5th Berkley Symp
  on Mathematical Statistics and Probability} \textbf{\bibinfo{volume}{1}},
  \bibinfo{pages}{666} (\bibinfo{year}{1967}).

\bibitem[{\citenamefont{Medus et~al.}(2005)\citenamefont{Medus, Acuna, and
  Dorso}}]{MedusA:Detcsn}
\bibinfo{author}{\bibfnamefont{A.}~\bibnamefont{Medus}},
  \bibinfo{author}{\bibfnamefont{G.}~\bibnamefont{Acuna}}, \bibnamefont{and}
  \bibinfo{author}{\bibfnamefont{C.~O.} \bibnamefont{Dorso}},
  \bibinfo{journal}{Physica A} \textbf{\bibinfo{volume}{358}},
  \bibinfo{pages}{593 } (\bibinfo{year}{2005}).

\bibitem[{\citenamefont{Holmstr\"{o}m et~al.}(2009)\citenamefont{Holmstr\"{o}m,
  Bock, and Br\"{a}nnlund}}]{Holmstrom20091161}
\bibinfo{author}{\bibfnamefont{E.}~\bibnamefont{Holmstr\"{o}m}},
  \bibinfo{author}{\bibfnamefont{N.}~\bibnamefont{Bock}}, \bibnamefont{and}
  \bibinfo{author}{\bibfnamefont{J.}~\bibnamefont{Br\"{a}nnlund}},
  \bibinfo{journal}{Physica D: Nonlinear Phenomena}
  \textbf{\bibinfo{volume}{238}}, \bibinfo{pages}{1161 }
  (\bibinfo{year}{2009}).

\bibitem[{Zac()}]{ZacharyURL}
\bibinfo{note}{\url{http://vlado.fmf.uni-lj.si/pub/networks/data/UciNet/UciDat%
a.htm}}.

\bibitem[{\citenamefont{Zachary}(1977)}]{Zachary77}
\bibinfo{author}{\bibfnamefont{W.~W.} \bibnamefont{Zachary}},
  \bibinfo{journal}{J. Anthropol. Res.} \textbf{\bibinfo{volume}{33}},
  \bibinfo{pages}{452} (\bibinfo{year}{1977}).

\bibitem[{\citenamefont{Bernard et~al.}(1980)\citenamefont{Bernard, Killworth,
  and Sailer}}]{Bernard80}
\bibinfo{author}{\bibfnamefont{H.}~\bibnamefont{Bernard}},
  \bibinfo{author}{\bibfnamefont{P.}~\bibnamefont{Killworth}},
  \bibnamefont{and} \bibinfo{author}{\bibfnamefont{L.}~\bibnamefont{Sailer}},
  \bibinfo{journal}{Social Networks} \textbf{\bibinfo{volume}{2}},
  \bibinfo{pages}{191} (\bibinfo{year}{1980}).

\bibitem[{\citenamefont{Bernard et~al.}(1982)\citenamefont{Bernard, Killworth,
  and Sailer}}]{Bernard82}
\bibinfo{author}{\bibfnamefont{H.}~\bibnamefont{Bernard}},
  \bibinfo{author}{\bibfnamefont{P.}~\bibnamefont{Killworth}},
  \bibnamefont{and} \bibinfo{author}{\bibfnamefont{L.}~\bibnamefont{Sailer}},
  \bibinfo{journal}{Social Science Research} \textbf{\bibinfo{volume}{11}},
  \bibinfo{pages}{30} (\bibinfo{year}{1982}).

\bibitem[{New()}]{NewmanURL}
\bibinfo{note}{\url{http://www-personal.umich.edu/~mejn/netdata/}}.

\bibitem[{\citenamefont{Lusseau et~al.}(2003)\citenamefont{Lusseau, Schneider,
  Boisseau, Haase, Slooten, and Dawson}}]{Lusseau2003}
\bibinfo{author}{\bibfnamefont{D.}~\bibnamefont{Lusseau}},
  \bibinfo{author}{\bibfnamefont{K.}~\bibnamefont{Schneider}},
  \bibinfo{author}{\bibfnamefont{O.~J.} \bibnamefont{Boisseau}},
  \bibinfo{author}{\bibfnamefont{P.}~\bibnamefont{Haase}},
  \bibinfo{author}{\bibfnamefont{E.}~\bibnamefont{Slooten}}, \bibnamefont{and}
  \bibinfo{author}{\bibfnamefont{S.~M.} \bibnamefont{Dawson}},
  \bibinfo{journal}{Behavioral Ecology and Sociobiology}
  \textbf{\bibinfo{volume}{54}}, \bibinfo{pages}{396} (\bibinfo{year}{2003}).

\bibitem[{\citenamefont{MacRae}(1960)}]{MacRae60}
\bibinfo{author}{\bibfnamefont{J.}~\bibnamefont{MacRae}},
  \bibinfo{journal}{Sociometry} \textbf{\bibinfo{volume}{23}},
  \bibinfo{pages}{360} (\bibinfo{year}{1960}).

\bibitem[{\citenamefont{Knuth}(1993)}]{Knuth93}
\bibinfo{author}{\bibfnamefont{D.~E.} \bibnamefont{Knuth}},
  \emph{\bibinfo{title}{A platform for combinatorial computing}},
  \bibinfo{howpublished}{The Stanford GraphBase, Addison-Wesley, Reading, MA}
  (\bibinfo{year}{1993}).

\bibitem[{Foo()}]{FoodwebURL}
\bibinfo{note}{\url{http://www.foodwebs.org/}}.

\bibitem[{Gle()}]{GleiserNote}
\bibinfo{note}{Data obtained through private communication}.

\bibitem[{\citenamefont{Guimera et~al.}(2003)\citenamefont{Guimera, Danon,
  Diaz-Guilera, Giralt, and Arenas}}]{Guimera2003}
\bibinfo{author}{\bibfnamefont{R.}~\bibnamefont{Guimera}},
  \bibinfo{author}{\bibfnamefont{L.}~\bibnamefont{Danon}},
  \bibinfo{author}{\bibfnamefont{A.}~\bibnamefont{Diaz-Guilera}},
  \bibinfo{author}{\bibfnamefont{F.}~\bibnamefont{Giralt}}, \bibnamefont{and}
  \bibinfo{author}{\bibfnamefont{A.}~\bibnamefont{Arenas}},
  \bibinfo{journal}{Phys. Rev. E} \textbf{\bibinfo{volume}{68}},
  \bibinfo{pages}{065103(R)} (\bibinfo{year}{2003}).

\bibitem[{Gui()}]{GuimeraNote}
\bibinfo{note}{Data obtained through private communication}.

\end{thebibliography}
\bibliographystyle{apsrev}

\end{document}